\documentclass[a4paper]{article}

\usepackage{INTERSPEECH2022}
\usepackage{multirow}
\usepackage{stfloats}
\usepackage{cite}
\usepackage{array}
\usepackage{float}
\usepackage{placeins}
\usepackage{varioref}

\title{Phase-Aware Spoof Speech Detection Based on Res2Net with Phase Network}
\name{Juntae Kim$^1$, Sung Min Ban$^1$}

\address{
  $^1$SK Telecom, Seoul, Republic of Korea}
\email{jtkim@kaist.ac.kr, sungmin.ban@sktelecom.com}

\begin{document}

\maketitle

\begin{abstract}
The spoof speech detection (SSD) is the essential countermeasure for automatic speaker verification systems. Although SSD with magnitude features in the frequency domain has shown promising results, the phase information also can be important to capture the artefacts of certain types of spoofing attacks. Thus, both magnitude and phase features must be considered to ensure the generalization ability to diverse types of spoofing attacks. In this paper, we investigate the failure reason of feature-level fusion of the previous works through the entropy analysis from which we found that the randomness difference between magnitude and phase features is large, which can interrupt the feature-level fusion via backend neural network; thus, we propose a phase network to reduce that difference. Our SSD system: phase network equipped Res2Net achieved significant performance improvement, specifically in the spoofing attack for which the phase information is considered to be important. Also, we demonstrate our SSD system in both known- and unknown-kind SSD scenarios for practical applications.

\end{abstract}

\section{Introduction}

In recent years, the automatic speaker verification (ASV) system\cite{kinnunen2010overview} has become an attractive research topic because ASV system has been widely applied into commercial applications. However, it has been known that ASV technologies are vulnerable to spoofing attacks. Thus, the countermeasure technologies for ASV: spoof speech detection (SSD) must be equipped, which discriminates between bona fide (genuine) audios and spoof ones. The spoofing attacks are mainly categorized into two types: physical access (PA) and logical access (LA). The PA considers the replay attack \cite{kinnunen2017asvspoof,you2020device,yang2020constant} while the LA includes the spoofing attacks based on text-to-speech synthesis\cite{zen2011product, wang2019vector,jang2021univnet,kim2018korean} and voice conversion \cite{tian2017exemplar,zhang2019sequence, sisman2019group} technologies.

Most of recent researches for SSD are based on ASVspoof2019\cite{wang2020asvspoof} in which both PA and LA are included, and the evaluation partition consists mostly of unseen spoofing attacks which differ from those in the training and development partitions. In this regard, the generalized ability to the various unseen spoofing attacks is the most important attribute for building the SSD system. 

We found that a number of researches have investigated various acoustic features to demonstrate their robustness to PA and LA with the state-of-the-art backend neural networks including: ResNet\cite{yang2021modified,yang2019sjtu,lai2019assert,jung2019replay,zhang122021effect}, Res2Net\cite{li2021replay,li2021channel}, and graph attention network\cite{jung2021aasist,tak2021end}. The representative acoustic features for SSD are: log-power (or magnitude) discrete Fourier transform (DFT), constant-Q transform (CQT), and linear frequency cepstral coefficients (LFCC), which are the magnitude features in the frequency domain\cite{li2021replay,jung2019replay, wang2021comparative,tak2020spoofing,9174990}. Among them, log-power CQT based approaches have shown outstanding performance both in PA and LA because CQT effectively represents low and high frequency bands with multiple frame windows with difference length, implying that CQT has fluent spectral information to capture the spoofing cues\cite{li2021replay}.

Although the approaches using magnitude features have shown promising results, there is a report that the use of the phase information can be important to detect the voice conversion based spoofing attack A17 known to be difficult to detect using magnitude\cite{takraw}. Considering the generalization ability to deal with various spoofing attacks, effectively utilizing both magnitude and phase features is an crucial research topic. 

In this work, the feature-level fusion between magnitude and phase features over Res2Net is achieved to boost the complementary effect of those features by adopting additional phase network reducing the randomness difference between magnitude and phase features prior to feeding them into Res2Net. To the best our knowledge, this is the first work showing promising results in LA, specifically for A17 via feature-level fusion while some works have demonstrated the benefits of score-level fusion. The contributions of this work also include: (i) investigating the failure reasons of the previous works for feature-level fusion through the entropy analysis; (ii) demonstrating the validity of the phase network both in known- and unknown-kind SSD scenarios.

\section{Proposed Method}
\subsection{Motivation}
There are some researches using phase information for SSD. In\cite{jung2021aasist}, the raw waveform is adopted to deploy the phase information with backend neural network named AAIST, showing extraordinary performance in LA, however, there is no demonstration in PA. In\cite{yang2021modified}, CQT with modified magnitude-phase spectrum (MMPS) was proposed to integrate the CQT-magnitude and -phase spectra in a hand-craft way. Although CQT-MMPS outperformed CQT-power spectrum in both PA and LA, there is no guarantee that the CQT-MMPS is an optimal method for combining the CQT-magnitude and -phase spectra for the synergy effect. Also, there is no investigation of A17 to demonstrate the effectiveness of phase information.

To boost the synergy effect between magnitude and phase spectra, it can be a choice that training a neural network with raw magnitude and phase spectra with a belief that neural network effectively utilizes the SSD related information from magnitude and phase spectra in a data-driven way. In\cite{jung2019replay}, the raw magnitude and phase spectra were concatenated across the channel dimension, and directly fed into the input convolution layer of ResNet to utilize the complementary phase information. Although individual magnitude and phase spectra were shown that they have respective discriminative power between bonafide and spoofed speech in PA, however, aggregating them did not show the synergy effect, even degrading the detection performance compared to the model using magnitude spectrum only. Further, there was no investigation of the effectiveness of phase information for LA. Due to the failure of feature-level fusion between magnitude and phase features, most of the combinations are conducted by developing separate systems for magnitude- and phase-based features and then fusing them at the score-level\cite{jelil2017spoof,das2018instantaneous}.

We assume that the failure of the neural network based deep fusion between magnitude and phase features is from the different feature complexity levels between magnitude and phase spectra. As the phase spectrum has inevitable wrapping problem\cite{vijayan2015analysis}, causing the randomness pattern; thus, analyzing some patterns from the phase spectrum is difficult than the magnitude spectrum which has well-structed patterns, e.g., harmonicity. 

To compare the randomness between magnitude and phase spectra, the entropy analysis was conducted to the CQT-magnitude and -phase spectra as described in Fig. 1. For the entropy analysis, the entropy is calculated per each time frame after global minmax normalization. We also analyze the entropy of random noise image to compare its randomness with that of CQT-phase spectrum. Note that each pixel of random noise image is independently sampled from the uniform random variable $X\sim U\left(0, 1\right)$. In Fig. 1, we found that CQT-phase spectrum and random noise image have similar entropy level, while the CQT-magnitude spectrum has the lower entropy level than the formers, specifically in the voice activity region (1.5--3.0 s), implying that CQT-phase spectrum has the higher randomness than the CQT-magnitude spectrum.

\begin{figure}[t]
  \centering
  \includegraphics[width=\linewidth]{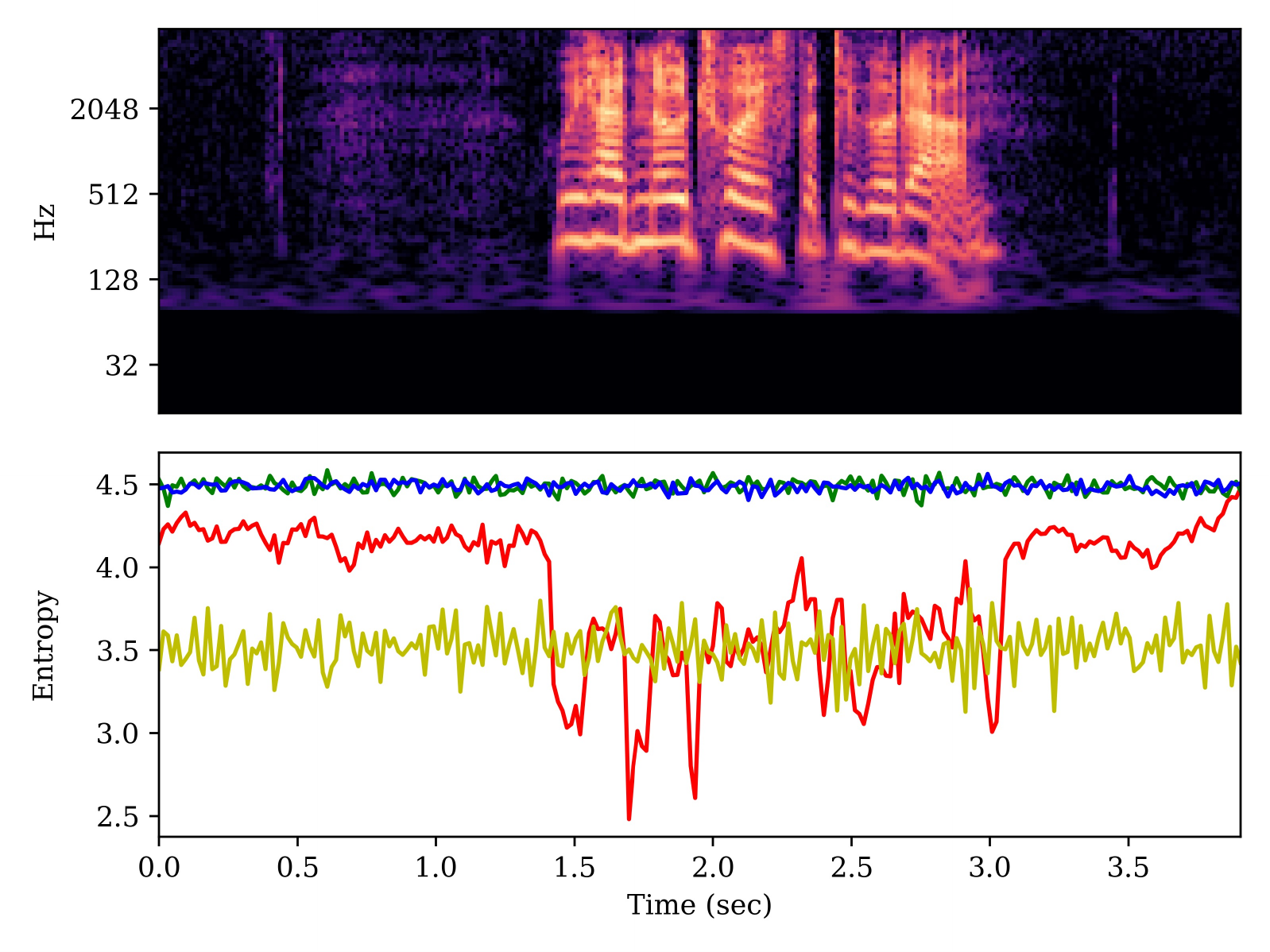}
  \caption{The CQT-magnitude spectrum in dB scale (up) and the entropy analysis (down) for the CQT-magnitude spectrum (red), CQT-phase spectrum (green), random noise image (blue), and the output features of phase network (yellow).}
\end{figure}

\begin{figure}[t]
  \centering
  \includegraphics[width=\linewidth]{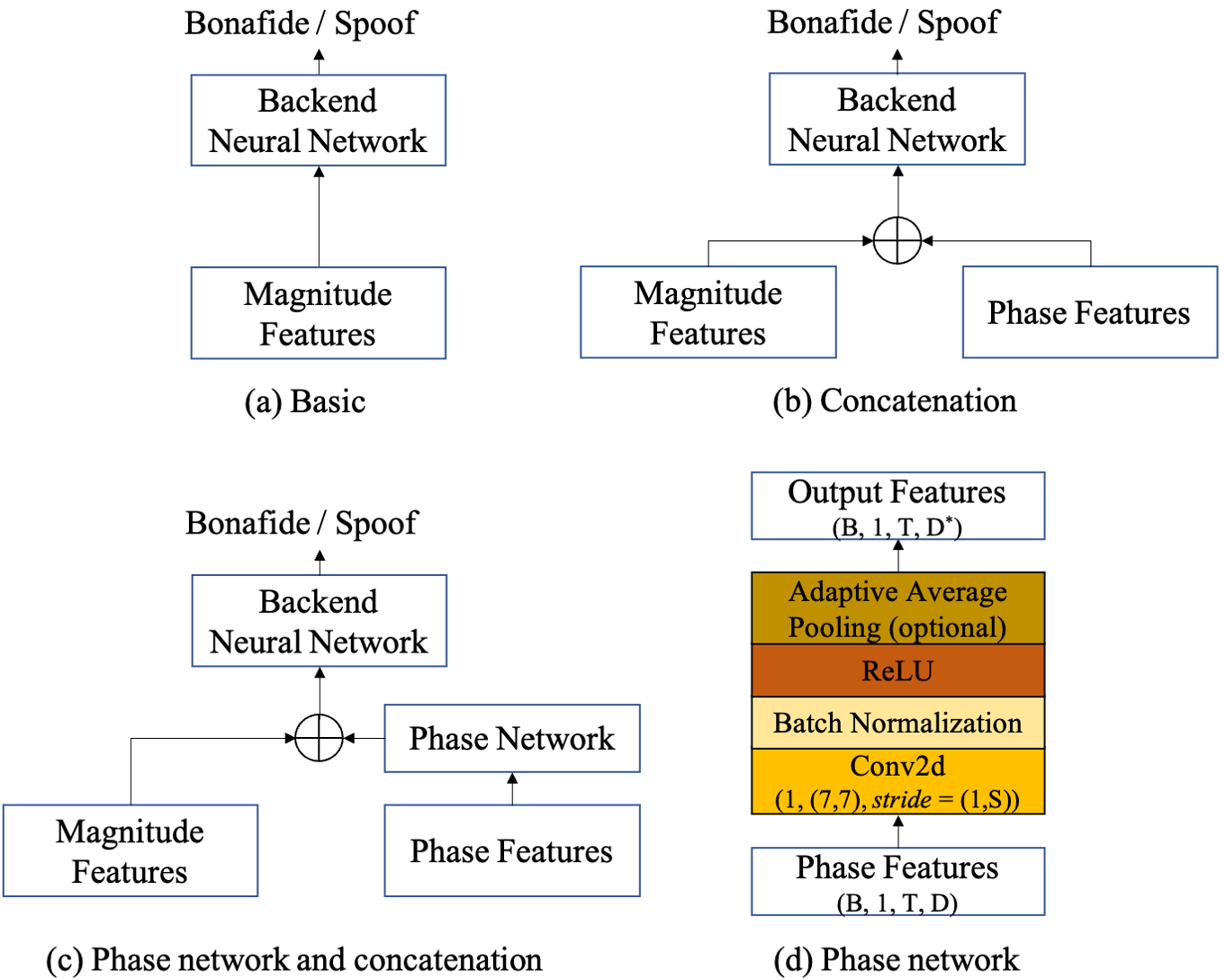}
  \caption{The investigated frameworks for SSD (a)--(c), and details of the phase network (d).}
\end{figure}

Due to the high randomness of phase spectrum, the training of the neural network with magnitude and phase spectra can be overwhelmed by the magnitude spectrum, i.e., highly depends on the magnitude spectrum, leading the additional model parameters for phase spectrum to be meaningless, suffering the model’s generalization ability known to important for SSD\cite{li2021replay}; thus, degrading the detection performance. This motivated us to propose the additional processing only for the phase spectrum to reduce its randomness for balancing the entropy level between magnitude and phase spectra before feeding them to the backend neural network model. Considering the RGB image based applications\cite{schwarz2015rgb, stiebel2018reconstructing} have shown promising results, and 3 channels in the RGB image have a similar randomness level each other, we can claim that our motivation to balance the randomness levels between magnitude and phase spectra is reasonable. 

The traditional method to reduce the randomness of phase spectrum is the phase unwrapping algorithms \cite{tribolet1977new,steiglitz1982phase,long1988exact} reconstructing the continuity of phase function, however, they have been known to be less accurate; thus, can distort the original phase information. In this regard, we propose the phase network to extract some useful information from the phase spectrum in a data-driven way. The details of phase network will be described in the next section. In Fig. 1, we also conducted the entropy analysis to the phase network’s output features, showing the similar entropy level compared to that of CQT-magnitude spectrum, and we believe that the phase features processed by the phase network instead of naïve CQT-phase spectrum can show the synergy effect with the CQT-magnitude spectrum because those features have almost similar entropy level; thus, the deep fusion can be achieved by backend neural network by effectively utilizing both CQT-phase and -magnitude spectra.
\subsection{Feature configuration}
Three types of well-known acoustic features for SSD are investigated in this work: CQT, DFT, and LFCC. For the magnitude features, the log-power operation is applied to CQT and DFT magnitude spectra; we followed the specification in\cite{li2021replay} for LFCC. To investigate the synergy effect between magnitude and phase spectra, the CQT and DFT log-power magnitude spectra are paired with CQT and DFT phase spectra, respectively; the LFCC is paired with DFT phase spectrum while the feature dimensions are different from each other. For the simplicity, we refer the log-power magnitude CQT and DFT spectra as CQT and LPS for the rest of the paper. 
\subsection{Frameworks for the phase-awareness}
To validate the effectiveness of using the phase information, we investigate three frameworks as described in Fig. 2 (a)--(c), where ‘$\oplus$’ implies the concatenation across the channel dimension. The framework (a) uses magnitude features alone, which is the most widely used framework for SSD. The framework (b) directly concatenates magnitude and phase features across the channel dimension. In the framework (c), the phase features firstly processed by phase network and the processed phase features and magnitude features are concatenated across the channel dimension. 


As the objective of the phase network is the feature extraction to reduce the randomness of original phase features, the phase network mainly consists of convolution layers. 
The details of the phase network is described in Fig. 2 (d), where B, T, D, and S is the batch size, number of time frames, feature dimension, and stride number. The $\text{D}^*$ is the feature dimension of magnitude features. For the convolution block (Conv2d), the first two numbers in parentheses indicate the channel size and kernel size, respectively. The adaptive average pooling layer is optionally applied when the feature dimensions between magnitude and phase features are different. In this work, $S=2$ (found from the development set) and average pooling were applied for LFCC paired with DFT phase; $S=1$ and no average pooling were applied for all other feature configurations. 

For backend neural network, Res2Net is adopted. Specifically, Res2Net50 with squeeze-and-excitation block introduced in\cite{li2021replay} is used as Res2Net with the same configuration. For simplicity, we refer Res2Net with framework (a), (b), and (c) in Fig. 2, as Res2Net-A, Res2Net-B, and Res2Net-C, respectively. Note that the input channel dimensions for Res2Net-A and Res2Net-B--C are 1 and 2, respectively. The number of model parameters for Res2Net-A--C is approximately 0.84M. As the phase network is the shallow convolution neural network, the parameter number difference between Res2Net-A and Res2Net-C is approximately 200, which is the minor number considering the total number of model parameters.

\begin{table*}[t]

\caption{The Results on ASVspoof2019 PA and LA in known-kind SSD scenario.  Performance reported in ``average(best)" after three repeated experiments.}
\label{table}
\setlength{\tabcolsep}{1pt}
\renewcommand{\arraystretch}{1.1}

\begin{tabular}{>{\centering}m{44pt}>{\centering}m{50.5pt}>{\centering}m{46.15pt}>{\centering}m{46.15pt}>{\centering}m{46.15pt}>{\centering}m{46.15pt}>{\centering}m{46.15pt}>{\centering}m{46.15pt}>{\centering}m{46.15pt}>{\centering}m{46.15pt}}
\noalign{\hrule height 0.8pt}

\multirow{ 3}{*}{\textbf{Model}} & \multirow{ 3}{*}{\textbf{Feature}} & \multicolumn{ 4}{c}{\textbf{PA}} & \multicolumn{ 4}{c}{\textbf{LA}}
\tabularnewline
\cline{3-10}
\multicolumn{ 2}{c}{} & \multicolumn{ 2}{c}{\textbf{Development}} & \multicolumn{ 2}{c}{\textbf{Evaluation}} & \multicolumn{ 2}{c}{\textbf{Development}} & \multicolumn{ 2}{c}{\textbf{Evaluation}} \tabularnewline
\cline{3-10}
\multicolumn{ 2}{c}{} & \textbf{t-DCF} & \textbf{EER(\%)} & \textbf{t-DCF} & \textbf{EER(\%)} & \textbf{t-DCF} & \textbf{EER(\%)} & \textbf{t-DCF} & \textbf{EER(\%)} \tabularnewline

\hline
Res2Net-A & LPS & 0.004(0.003) &	0.15(0.13) &	0.009(0.007) &	0.35(0.28) &	0.000(0.000) &	0.00(0.00) &	0.283(0.194) &	14.0(11.1) \tabularnewline
Res2Net-B & LPS+phase&0.005(0.004)  &	0.17(0.15) & 0.008(0.007) &	0.31(0.27) &	0.000(0.000) &	0.00(0.00) &	0.309(0.280) &	12.6(12.0) \tabularnewline
Res2Net-C & LPS+phase&0.005(0.004)  &	0.16(0.15) &	0.008(0.008) &	0.34(0.31) &	0.000(0.000) &	0.00(0.00) &	0.225(0.185) &	10.8(8.08) \tabularnewline

\hline
Res2Net-A & LFCC & 0.021(0.020) &	0.86(0.85) &	0.042(0.039) &	1.46(1.37) &	0.005(0.005) &	0.18(0.15) &	0.062(0.049) &	2.37(1.81) \tabularnewline
Res2Net-B & LFCC+phase & 0.018(0.016) &	0.71(0.67) &	0.037(0.033) &	1.26(1.14) &	0.005(0.003) &	0.19(0.16) &	0.091(0.069) &	3.48(2.58) \tabularnewline
Res2Net-C & LFCC+phase & 0.057(0.056)	& 2.06(1.97) &	0.082(0.077) &	2.93(2.76) &	0.004(0.003) &	0.18(0.12) &	0.056(0.048) &	2.09(1.67) \tabularnewline
\hline

Res2Net-A & CQT & 0.007(0.005) &	0.26(0.18) &	0.016(0.014) &	0.58(0.51) &	0.006(0.004) &	0.23(0.12) &	0.050(0.038) &	1.70(1.31) \tabularnewline
Res2Net-B & CQT+phase &0.009(0.006) &	0.34(0.24) &	0.017(0.017) &	0.65(0.63) &	0.007(0.005) &	0.24(0.19) &	0.052(0.046) &	1.76(1.47) \tabularnewline
\textbf{Res2Net-C} & \textbf{CQT+phase} & \textbf{0.007(0.006)} &	\textbf{0.28(0.22)} &	\textbf{0.017(0.015)} &	\textbf{0.63(0.56)} &	\textbf{0.006(0.005)} &	\textbf{0.20(0.16)} &	\textbf{0.030(0.027)} &	\textbf{1.07(0.94)} \tabularnewline
\hline

Res2Net-A & CQT-MMPS & 0.007\,(0.004) &	0.26(0.19) &	0.014(0.012) &	0.51(0.48) &	0.026(0.022) &	0.82(0.70) &	0.137(0.109) &	4.84(3.64) \tabularnewline
AAIST & waveform & 0.817\,(0.602) &	36.3(23.1) &	0.814(0.685) &	34.4(25.8) &	0.016(0.014) &	0.54(0.51) &	0.043(0.038) &	1.45(1.29) \tabularnewline
\hline
\noalign{\hrule height 0.8pt}
\end{tabular}
\label{tab3}
\end{table*}

\section{Experiments and Results}
\subsection{Experimental setup}
\textbf{Dataset:} All experiments are conducted on the ASVspoof 2019 challenge dataset, which contains two subsets: PA and LA. Each subset includes training, development, and evaluation partitions. The training and development partitions are used for model training and selection, respectively, and the evaluation partition is used for model evaluation. The further details for the used dataset can be found in\cite{wang2020asvspoof}.

\textbf{Evaluation metrics:} The trained models are evaluated by the tandem detection cost function (t-DCF)\cite{todisco2019asvspoof} and equal error rate (EER)\cite{todisco2019asvspoof}. The log-probability of the bonafide class is used as the score for t-DCF and EER computation.

\textbf{Feature specification:} The CQT is extracted with 16ms hop length, Hanning window, and 9 octaves with 12 bins per octave; thus, the feature dimension for CQT is 108. The DFT is extracted with a Hanning window; the window and hop lengths are 64ms and 32ms, respectively, and 1024 FFT points are used; thus, the feature dimension for DFT is 513. As in\cite{li2021replay}, LFCC is extracted with 20ms window length and 10ms hop length with 512 FFT points and 20 filters; the delta and delta-delta features are also included; thus, the feature dimension for LFCC is 60.

\textbf{Feature engineering:} As the lengths of speech utterances are different from each other, some feature engineering is necessary. In this work, the unified feature map method in\cite{lai2019assert} is adopted. Therefore, we first extended all utterances to multiple of 400 frames. After that, the extended utterance was broken down into segments of length 400 frames with 200 frames overlap. For evaluation, the backend neural network outputs from multiple segments were simply averaged for each utterance.

\textbf{Training and evaluation scenario:} We found that most of SSD researches follow the known-kind SSD scenario, e.g., if the model is trained on PA, the model is only evaluated in PA. However, the practical scenario is the unknown-kind SSD scenario in which the model is trained on both PA and LA to detect the spoof speech regardless of the spoof type. This work includes both known- and unknown-kind SSD scenarios to validate the proposed method in practical environment.

\textbf{Training strategy:} The training strategy is similar with\cite{li2021replay}. The binary cross entropy loss is used to train all models. Adam is adopted as the optimizer with $\beta_{1}=0.9$, $\beta_{2}=0.98$ and weight decay $10^{-5}$. All models are trained with 30 epochs, and the model with lowest EER on development set is chosen to be evaluated. Further, we found that the SSD performance using ASVspoof 2019 has high variation according to the initialization weights\cite{wang2021comparative}. To strictly validate the proposed method, all experiments are repeated by three times with different random seeds.  

\subsection{Experimental results and discussion}
\begin{table*}[t]

\label{table}
\setlength{\tabcolsep}{1pt}
\renewcommand{\arraystretch}{1.1}

\caption{Breakdown EER (\%) performance of all 13 attacks from ASVspoof2019 LA-Evaluation, pooled min t-DCF(P1) and pooled EER(\%, P2). The reported performances are from the best performed models shown in Table 1.}
\begin{tabular}{>{\centering}m{44pt}>{\centering}m{50.5pt}>{\centering}m{23pt}>{\centering}m{23pt}>{\centering}m{23pt}>{\centering}m{23pt}>{\centering}m{23pt}>{\centering}m{23pt}>{\centering}m{23pt}>{\centering}m{23pt}>{\centering}m{23pt}>{\centering}m{23pt}>{\centering}m{23pt}>{\centering}m{23pt}>{\centering}m{23pt}>{\centering}m{28.3pt}>{\centering}m{28.3pt}}
\noalign{\hrule height 0.8pt}

\textbf{Model} & \textbf{Feature} & \textbf{A07} &	\textbf{A08} &	\textbf{A09} &	\textbf{A10} &	\textbf{A11} &	\textbf{A12} &	\textbf{A13} &	\textbf{A14} & \textbf{A15} & \textbf{A16} & \textbf{A17} & \textbf{A18} & \textbf{A19} & \textbf{P1} & \textbf{P2} \tabularnewline
\hline
Res2Net-A & CQT & 0.16 &	1.73 &	0.02 &	0.51 &	0.24 &	0.10 &	0.12 &	0.19 & 0.47 & 0.55 & 2.99 & 2.62 & 0.88 & 0.038 & 1.31 \tabularnewline
Res2Net-B & CQT+phase & 0.19 &	0.90 &	0.02 &	0.41 &	0.22 &	0.11 &	0.12 &	0.16 & 0.20 & 0.22 & 3.82 & 2.60 & 1.16 & 0.046 & 1.47 \tabularnewline

\textbf{Res2Net-C} & \textbf{CQT+phase} & \textbf{0.22} &	\textbf{2.67} &	\textbf{0.02} &	\textbf{0.51} &	\textbf{0.33} &	\textbf{0.18} &	\textbf{0.06} &	\textbf{0.22} & \textbf{0.22} & \textbf{0.34} & \textbf{1.75} & \textbf{1.77} & \textbf{0.33} & \textbf{0.027} & \textbf{0.94} \tabularnewline

\hline
\noalign{\hrule height 0.8pt}
\end{tabular}

\label{tab4}
\medskip
\caption{The Results on ASVspoof2019 PA and LA in unknown-kind SSD scenario.  Performance reported in ``average(best)" after three repeated experiments.}

\begin{tabular}{>{\centering}m{44pt}>{\centering}m{50.5pt}>{\centering}m{46.15pt}>{\centering}m{46.15pt}>{\centering}m{46.15pt}>{\centering}m{46.15pt}>{\centering}m{46.15pt}>{\centering}m{46.15pt}>{\centering}m{46.15pt}>{\centering}m{46.15pt}}
\noalign{\hrule height 0.8pt}

\multirow{ 3}{*}{\textbf{Model}} & \multirow{ 3}{*}{\textbf{Feature}} & \multicolumn{ 4}{c}{\textbf{PA}} & \multicolumn{ 4}{c}{\textbf{LA}}
\tabularnewline
\cline{3-10}
\multicolumn{ 2}{c}{} & \multicolumn{ 2}{c}{\textbf{Development}} & \multicolumn{ 2}{c}{\textbf{Evaluation}} & \multicolumn{ 2}{c}{\textbf{Development}} & \multicolumn{ 2}{c}{\textbf{Evaluation}} \tabularnewline
\cline{3-10}
\multicolumn{ 2}{c}{} & \textbf{t-DCF} & \textbf{EER(\%)} & \textbf{t-DCF} & \textbf{EER(\%)} & \textbf{t-DCF} & \textbf{EER(\%)} & \textbf{t-DCF} & \textbf{EER(\%)} \tabularnewline

\hline

Res2Net-A & CQT & 0.009(0.007) &	0.35(0.28) &	0.016(0.012) &	0.61(0.45) &	0.015(0.013) &	0.46(0.40) &	0.111(0.097) &	3.84(3.66) \tabularnewline
Res2Net-B & CQT+phase &0.011(0.009) &	0.45(0.30) &	0.019(0.016) &	0.69(0.54) &	0.009(0.008) &	0.31(0.27) &	0.112(0.105) &	4.54(4.09) \tabularnewline
\textbf{Res2Net-C} & \textbf{CQT+phase} & \textbf{0.009(0.008)} &	\textbf{0.39(0.31)} &	\textbf{0.016(0.013)} &	\textbf{0.58(0.49)} &	\textbf{0.012(0.007)} &	\textbf{0.38(0.23)} &	\textbf{0.093(0.092)} &	\textbf{3.33(3.23)} \tabularnewline
\hline

Res2Net-A & CQT-MMPS & 0.014(0.008) &	0.53(0.31) &	0.018(0.010) &	0.66(0.40) &	0.036(0.034) &	1.16(1.09) &	0.147(0.128) &	6.27(5.51) \tabularnewline

\hline
\noalign{\hrule height 0.8pt}
\end{tabular}

\end{table*}

For known-kind SSD scenario, Table 1 compares Res2Net-A--C with different acoustic feature configurations; Res2Net with CQT-MMPS and AAIST are also compared for the baselines. We directly implemented the CQT-MMPS and used the open source toolkit for AAIST\footnote{ https://github.com/clovaai/aasist}. In LA-Evaluation, we found that Res2Net-C with CQT+phase outperformed all other methods in a great margin in both t-DCF and EER. In contrast, the Res2Net-B showed the worse performance than Res2Net-A in all feature configurations, excepting the averaged EER comparison in LPS configuration. These results implies that simply concatenated phase and magnitude features did not show any synergy effect, i.e., the Res2Net could not extract any useful information from the phase features for the synergy effect, degrading the performance due to additional  meaningless parameters deteriorating the model's generalization ability. In this regard, we can claim that the phase network must be leveraged to effectively deploy the phase information for synergy effect between magnitude and phase features. Among the baselines, AAIST showed promising performance as in\cite{jung2021aasist}, while Res2Net-A with CQT-MMPS showed the inferior performance.

In LA-Development, Res2Net-A--C with LPS feature configuration showed the best performance; however, it can be the results from the overfitting to known spoofing types because Res2Net-A--C with LPS feature configuration showed the worst performance in LA-Evaluation. Res2Net-A--C with LFCC and CQT configurations showed comparable performance each other, outperforming AAIST and Res2Net-A with CQT-MMPS.

In PA-Evaluation and -Development, Res2Net-A--C with LPS feature configuration showed the best performance; however, there is no significant performance improvement between Res2Net-A and Res2Net-C, implying the phase information for PA is less important than LA. Also, we did not investigate the performance degradation between Res2Net-A and Res2Net-B with LPS feature configuration, which is shown in\cite{jung2019replay}. As the main difference is using the Res2Net for backend neural network instead of ResNet used in \cite{jung2019replay}, we can consider that Res2Net is more robust than ResNet to the overfitting problem for additional parameters used to process the phase-related part. Res2Net-A--C with CQT feature configuration showed the reasonable performance. Among the baselines, AAIST showed the worst performance, while Res2Net-A with CQT-MMPS showed outstanding performance. In this regard, AAIST and Res2Net-A with CQT-MMPS are considered to biased frameworks to LA and PA, respectively. Considering the performance both in LA and PA, we choose the Res2Net-A--C with CQT feature configuration for unknown-kind SSD. 

To validate whether Res2Net-C with CQT+phase effectively utilizes the phase features, we investigated the best performed models in Table 2: Res2Net-A--C with CQT feature configuration, according to the 13 attack types for LA-Evaluation as described in Table 2. In \cite{takraw}, the phase information is considered to be important for detecting the voice conversion attack A17 in Table 2; thus, the performance for A17 attack can be an indicator whether the model effectively exploits the phase information. In Table 2, we found that Res2Net-C significantly outperformed Res2Net-A for A17 (relatively 41.4\% improvement), while the Res2Net-B showed the inferior performance than Res2Net-A; thus, we can claim that the proposed Res2Net-C with CQT+phase effectively utilized the phase information, showing the synergy effect with magnitude features. Also, we found that ResNet-A and -B performed better than Res2Net-C in some attacks, however, Res2Net-C still showed the comparable EER ($<1\%$), excepting for A08. We found that the magnitude features are important for A08\cite{zhang122021effect}; thus, mixing some phase-related information can degrade the performance.

For unknown-kind SSD scenario, Table 3 compares Res2Net-A--C with CQT feature configuration including CQT-MMPS because CQT-MMPS based model showed comparable performance in this scenario\cite{yang2021modified}. We found that Res2Net-C with CQT+phase outperformed all other methods in LA-Evaluation and -Development, while showing the comparable performance compared to Res2Net-A with CQT in PA-Evaluation and -Development; thus, we claim that Res2Net-C with CQT+phase is appropriate for the practical SSD application. The Res2Net-B showed the inferior performance than Res2Net-A with CQT and Res2Net-C with CQT+phase in both PA and LA.

\section{Conclusion}

We propose the phase-aware SSD based on Res2Net with phase network. To obtain the synergy effect between magnitude and phase features, the phase network is adopted for individual processing for phase features to reduce their intrinsic high randomness. Although proposed method showed outstanding performance for LA in known-kind-scenario, there is a performance degradation for LA in unknown-kind-scenario; thus, reconstructing that performance is our future work.

\bibliographystyle{IEEEtran}

\bibliography{mybib}

\end{document}